\documentclass[prd,twocolumn,floatfix,preprintnumbers,nofootinbib,nopacs]{revtex4}
\usepackage[utf8x]{inputenc}

\usepackage{amsmath,amssymb,amsfonts,mathrsfs,bbold}

\usepackage{graphicx}

\usepackage[caption = false]{subfig}
\usepackage[normalem]{ulem}
\usepackage{xcolor}
\definecolor{lcolor}{rgb}{0.5,0,0}
\definecolor{citcolor}{rgb}{0,0.3,0.0}
\usepackage{nicefrac}
\usepackage{physics}
\usepackage{siunitx}

\usepackage[breaklinks,colorlinks,urlcolor=blue,citecolor=citcolor,linkcolor=lcolor]{hyperref}
\usepackage{mciteplus}

\usepackage{physics}

\newcommand{\nc}{{N_\mathrm{c}}}
\newcommand{\nf}{{N_\mathrm{f}}}

\newcommand{\as}{\alpha_{\mathrm{s}}}

\begin{document}

\author{Daniel Adamiak}
    \email[Email: ]{dadamiak@jlab.org}
        \affiliation{Jefferson Lab, Newport News, VA 23606, USA}

\author{Heikki Mäntysaari}
\email[Email: ]{heikki.mantysaari@jyu.fi}
\affiliation{
Department of Physics, University of Jyväskylä,  P.O. Box 35, 40014 University of Jyväskylä, Finland
}
\affiliation{
Helsinki Institute of Physics, P.O. Box 64, 00014 University of Helsinki, Finland
}

\author{Yossathorn Tawabutr}
    \email[Email: ]{yossathorn.t@chula.ac.th}
    \affiliation{
Department of Physics, University of Jyv\"askyl\"a,  P.O. Box 35, 40014 University of Jyv\"askyl\"a, Finland
}
    \affiliation{
Helsinki Institute of Physics, P.O. Box 64, 00014 University of Helsinki, Finland
}
\affiliation{Department of Physics, Faculty of Science, Chulalongkorn University, 254 Phaya Thai Rd, Wang Mai, Pathum Wan, Bangkok 10330, Thailand}

\title{Proton spin from small-$x$ with constraints from the valence quark model}

\begin{abstract}
    We apply the valence quark model~\cite{Dumitru:2024pcv} to constrain the non-perturbative initial condition for the small-$x$ helicity evolution. The remaining free parameters are constrained by performing a global analysis akin to~\cite{Adamiak:2023yhz} to the available polarized small-$x$ deep inelastic scattering data. A good description of the world data is obtained with only 8 free parameters. The model parameters are tightly constrained by the data, allowing us to predict the proton polarized structure function $g_1^p$ to be negative at small $x$. Furthermore, we obtain the small-$x$ quark and gluon spins to give a contribution
   $ \int_{10^{-5}}^{0.1} \dd x \left( \frac{1}{2}\Delta\Sigma + \Delta G \right) = 0.63 \pm 0.10$ or $1.35\pm 0.16$
   to the proton spin, depending on the applied running coupling prescription.
\end{abstract}

\preprint{JLAB-THY-25-4239}

\maketitle

\section{Introduction}


The proton spin puzzle, concerning the amount of contribution to the proton spin that comes from the spin and orbital angular momenta of quarks and gluons inside, is an actively studied problem in contemporary particle physics~\cite{Bartels:1996wc,Leader:2013jra,Kovchegov:2015pbl,Kovchegov:2016zex,Kovchegov:2017lsr,Kovchegov:2018znm,Kovchegov:2020hgb,Cougoulic:2020tbc,Adamiak:2021ppq,Kovchegov:2021lvz,Kovchegov:2021iyc,Cougoulic:2022gbk,Kovchegov:2022kyy,Adamiak:2023okq,Borden:2023ugd,Adamiak:2023yhz,Kovchegov:2023yzd,Kovchegov:2024aus,Borden:2024bxa,Kovchegov:2024wjs,Chirilli:2018kkw,Chirilli:2021lif,Bhattacharya:2022vvo,Bhattacharya:2024sno,Aschenauer:2013woa,Aschenauer:2015eha,Dumitru:2024pcv}. This letter focuses on the quark and gluon spin contributions, which can be written as the integrals of helicity-dependent parton distribution functions (hPDFs) over the whole range of parton's longitudinal momentum fraction, $x$. 
Experimentally accessing hPDFs at small-$x$  requires polarized scattering processes at large center-of-mass energy, which are costly to achieve. Consequently, experimental data are only available down to $x_{\min}=0.001$ for quarks and $x_{\min}=0.05$ for gluons~\cite{Aschenauer:2013woa,Aschenauer:2015eha}. 

In order to complement the available experimental results and to determine the total quark and gluon contribution to the proton spin, a high-energy evolution equation has been derived~\cite{Kovchegov:2015pbl,Kovchegov:2016zex,Kovchegov:2018znm,Cougoulic:2022gbk} for hPDFs. This evolution equation allows for hPDFs at lower $x$'s to be computed from the counterparts at moderate $x$'s, which can be determined from experimental measurements. The helicity evolution equation is similar in construction to its counterpart for unpolarized PDFs -- the Balitsky-Kovchegov (BK) equation~\cite{Balitsky:1995ub,Kovchegov:1999yj} -- in that it employs as degrees of freedom the \emph{polarized dipole amplitudes,} which are the amplitudes of the polarized scattering between the target and a $q\bar{q}$ dipole within the dipole framework of deep-inelastic scattering (DIS)~\cite{Mueller:1989st,Nikolaev:1990ja}. However, in contrast to the BK equation that resums $\as\ln(1/x)$ per evolution step, the helicity evolution equation resums $\as\ln^2(1/x)$ per iteration, that is, it is within the double-logarithmic (DLA) level that dominates the leading-logarithmic (LLA) regime of the BK equation at small $x$. Here, $\as$ is the strong coupling constant.
 
Helicity-dependent scattering processes at high center-of-mass energy receive equally significant contributions from quark and gluon exchanges with the target, in contrast to the unpolarized counterparts for which gluon exchange dominates~\cite{Kovchegov:2015pbl,Kovchegov:2018znm,Kovchegov:2021iyc}. 
As a result, the polarized dipole amplitudes vary with the flavor of the $q\bar{q}$ pair. This leads to the C-even dipole amplitudes, $Q_f(r_{\perp},zs)$, corresponding to the C-even contributions to the amplitude of a $q\bar{q}$ pair of flavor $f\in\{u,d,s\}$ and transverse size $r_{\perp}$, with the polarized (anti)quark having longitudinal momentum fraction $z$, interacting with the target. Here, $s$ in the variable, $zs$, is the squared center-of-mass energy of the dipole-target scattering. Similarly, we also have the C-odd counterparts, $Q_f^{\text{NS}}(r_{\perp},zs)$. Furthermore, in the case where the $q\bar{q}$ dipole is in fact half of a gluon dipole in the limit where $N_\mathrm{c}$ -- the number of quark colors -- is large~\cite{tHooft:1973alw}, the polarized scattering amplitude is denoted by ${\widetilde G}(r_{\perp},zs)$. Finally, the evolution equation of all these polarized dipole amplitudes depends on an extra amplitude, $G_2(r_{\perp},zs)$, which describes the helicity-independent gluon exchange that likely relates to the strong-interaction counterpart of spin-orbit coupling~\cite{Cougoulic:2022gbk}. Altogether, the small-$x$ helicity evolution equation involves eight polarized dipole amplitudes in the Veneziano large-$\nc\& \nf$ limit~\cite{Veneziano:1976wm}\footnote{See~\cite{Adamiak:2023yhz} for a more detailed description of the 8 dipole amplitudes introduced here.}. In this limit, the evolution equation becomes closed and linear, allowing for iterative solutions to be obtained via discretization of the transverse and rapidity space for any given moderate-$x$ initial conditions. 

The hPDFs $\Delta \Sigma$ and $\Delta G$, describing the quark and gluon contributions to the proton spin, together with the relevant $g_1$ structure function, can be computed from the polarized dipole amplitudes via the following formulas derived in~\cite{Cougoulic:2022gbk}:
\begin{subequations}\label{hPDF_g1}
\begin{align}
    \Delta\Sigma(x,Q^2) &= -\frac{\nc}{2\pi^3}\sum_f\int\limits_{\Lambda^2/s}^1\frac{\dd z}{z}\int\limits_{1/zs}^{\min\left[\frac{1}{zQ^2},\frac{1}{\Lambda^2}\right]}\frac{\dd r^2_{\perp}}{r^2_{\perp}}  \label{qk_hPDF} \\
    &\;\;\;\;\times \left[Q_f(r_{\perp},zs) + 2G_2(r_{\perp},zs)\right]  , \notag \\
    \Delta G(x,Q^2) &= \frac{2\nc}{\as\pi^2} \,G_2\left(r_{\perp}=\frac{1}{Q},zs=\frac{Q^2}{x}\right)  , \label{gl_hPDF} \\
    g_1(x,Q^2) &= - \frac{\nc}{4\pi^3}\sum_fe^2_f \int\limits_{\Lambda^2/s}^1\frac{\dd z}{z}\int\limits_{1/zs}^{\min\left[\frac{1}{zQ^2},\frac{1}{\Lambda^2}\right]}\frac{\dd r^2_{\perp}}{r^2_{\perp}}  \label{g1} \\
    &\;\;\;\;\times \left[Q_f(r_{\perp},zs) + 2G_2(r_{\perp},zs)\right]  . \notag
\end{align}
\end{subequations}
This allows for the hPDFs at small $x$ to be computed from the polarized dipole amplitudes evolved to large evolution rapidities (small $x$), that is, the high-energy evolution is run solely in terms of the dipole amplitudes. Here $\Lambda$ is an infrared regulator, and following \cite{Adamiak:2023yhz} we set $\Lambda=\SI{1}{GeV}$.

In order to compute helicity PDFs at small-$x$, the perturbative helicity evolution equations have to be combined with non-perturbative initial conditions that can be extracted from data.
This letter presents a new fit with the valence-quark-model constrained initial conditions~\cite{Dumitru:2024pcv} to the small-$x$ helicity evolution at large $\nc\& \nf$~\cite{Cougoulic:2022gbk}, together with its phenomenological consequences. Section~\ref{sec:setup} specifies the exact setup of the initial conditions and the fit, together with the running coupling prescription. Section~\ref{sec:results} details the results and predictions to quark and gluon hPDFs, $g_1$ structure functions and other physical predictions of interest. Finally, Section~\ref{sec:conclusion} provides the conclusion and potential directions for future works.

\section{Computing helicity PDFs at small $x$}
\label{sec:setup}

An essential ingredient in determining the hPDFs is the set of initial conditions to the evolution equation of the polarized dipole amplitudes. These initial conditions correspond to scatterings at moderate center-of-mass energy and hence can be deduced from experimental data. The standard process is to begin with a model that contains a number of unknown parameters, run the evolution equation, and then compare the resulting small-$x$ hPDFs to the available data. The quality of such fits depends on the model choice and the amount of available data. A recent global analysis~\cite{Adamiak:2023yhz} employs the model inspired by the Born-level expression~\cite{Kovchegov:2015pbl,Kovchegov:2016zex}, writing each dipole amplitude as a linear combination of the logarithm, $\ln(1/r^2_{\perp}\Lambda^2)$, of the transverse dipole size, the logarithm, $\ln(zs/\Lambda^2)$, of the longitudinal momentum fraction, together with a constant term. As such, the initial condition is parametrized as
\begin{subequations}\label{Born_IC}
\begin{align}
    Q_f^{(0)}(r_{\perp},zs) &= a_f\ln\frac{zs}{\Lambda^2} + b_f\ln\frac{1}{r^2_{\perp}\Lambda^2} + c_f \, , \\
    Q_f^{\text{NS}(0)}(r_{\perp},zs) &= a_f^{\text{NS}}\ln\frac{zs}{\Lambda^2} + b_f^{\text{NS}}\ln\frac{1}{r^2_{\perp}\Lambda^2} + c_f^{\text{NS}} \, , \\
    {\widetilde G}(r_{\perp},zs) &= {\widetilde a}\ln\frac{zs}{\Lambda^2} + {\widetilde b}\ln\frac{1}{r^2_{\perp}\Lambda^2} + {\widetilde c} \, , \\
    G_2(r_{\perp},zs) &= a_2\ln\frac{zs}{\Lambda^2} + b_2\ln\frac{1}{r^2_{\perp}\Lambda^2} + c_2 \, .
\end{align}
\end{subequations}
With the three lightest flavors, $f\in\{u,d,s\}$, this model contains 24 free parameters. In~\cite{Adamiak:2023yhz}, the dipole amplitudes from this initial condition are evolved to large evolution rapidities and employed to compute the asymmetries $A_1$, $A_{\parallel}$ and $A_1^h$ for polarized DIS and SIDIS processes. Here, $h$ denotes the produced hadron in SIDIS processes. The unpolarized cross-section needed for these asymmetry observables is taken from a previous JAM analysis and incorporated as in \cite{Adamiak:2023yhz}. The results are then compared to the 226 available data points within $0.005\leq x\leq 0.1$ and $1.69\text{ GeV}^2\leq Q^2\leq 10.4\text{ GeV}^2$ from SLAC~\cite{E142:1996thl,E143:1998hbs,E154:1997xfa,E155:1999pwm,E155:2000qdr}, EMC~\cite{EuropeanMuon:1989yki}, SMC~\cite{SpinMuon:1997yns,SpinMuon:1998eqa,SpinMuon:1999udj}, COMPASS~\cite{COMPASS:2009kiy,COMPASS:2010hwr,COMPASS:2010wkz,COMPASS:2015mhb,COMPASS:2016jwv} and HERMES~\cite{HERMES:1997hjr,HERMES:1999uyx,HERMES:2004zsh,HERMES:2006jyl} for the proton, deuterium and helium-3 targets. Overall, this leads to an excellent fit with $\chi^2=1.03$ per degrees of freedom. This implies the consistency between the small-$x$ helicity evolution equation from~\cite{Cougoulic:2022gbk} and the available polarized scattering data. However, when it comes to predictions, there remains significant variation among the replicas. Consequently it is not possible to obtain solid answers to central questions like the small-$x$ gluon contribution to the proton spin, and the sign of the $g_1$ structure function at small $x$~\cite{Adamiak:2023yhz}.
This is likely a sign that the initial condition employed contains a large number of free parameters compared to the amount of available data.

In~\cite{Dumitru:2024pcv}, a calculation is performed using the valence quark model of the proton target with the inclusion of one perturbative gluon emission in order to determine the polarized dipole amplitudes at moderate $x$. The operator forms of all the quark- and gluon-exchange terms in the definitions of various polarized dipole amplitudes are computed within the valence quark framework previously applied in the context of unpolarized scattering e.g. in Refs.~\cite{Dumitru:2023sjd,Dumitru:2021tvw,Dumitru:2020gla,Dumitru:2018vpr}. As a result, upon putting the squared center-of-mass energy, $zs$, as the hardest scale of the problem, we are able to fix 16 of the 24 parameters in Eqs.~\eqref{Born_IC} up to the coupling constant, $\as$, leaving only 8 parameters to be determined through the fit. Thus, we expect the physical input from the valence quark model to provide a sufficiently strict model for the initial conditions, so that meaningful predictions can be obtained from a new fit to the available polarized DIS and SIDIS measurements at small $x$. 

The valence quark model of Ref.~\cite{Dumitru:2024pcv} is only applicable at moderatetely large $x$. This in principle is a challenge especially when the valence quark model initial condition is used as an input for the helicity-independent BK evolution valid at small-$x$ resumming contributions $\as \ln 1/x$~\cite{Dumitru:2023sjd}. On the other hand, the helicity evolution equation~\cite{Kovchegov:2015pbl,Kovchegov:2016zex,Kovchegov:2018znm,Cougoulic:2022gbk} resums contributions $\as \ln^2 1/x$ to all orders, and consequently it is expected to be accurate at much higher values of Bjorken-$x$ than its unpolarized counterpart. As such, we expect the valence quark model to provide a realistic initial condition at $x=x_0=0.1$ in this work.







The valence quark model calculation~\cite{Dumitru:2024pcv} fixes all 16 parameters multiplying logarithms in Eqs.~\eqref{Born_IC}  to 
\begin{subequations}\label{val_qk_params}
\begin{align}
    a_f &= \frac{4\pi\as^2}{81}\left[\delta_{f,u}+11\delta_{f,d}+9\delta_{f,s}\right] \, \overline{x^{-1}} \,, \label{val_qk_af} \\
    a_f^{\text{NS}} &= -\frac{8\pi\as^2}{81}\left[4\delta_{f,u}-\delta_{f,d} \right] \, \overline{x^{-1}} \,, \label{val_qk_afNS} \\
    {\widetilde a} &= \frac{23\pi\as^2}{18} \, \overline{x^{-1}} \,, \label{val_qk_at} \\
    a_2 &= 0\,,\label{val_qk_a2} \\
    b_f &= - \frac{2\pi\as^2}{81}\left[18+4\delta_{f,u}-\delta_{f,d}\right] \, \overline{x^{-1}} \,, \label{val_qk_bf} \\
    b_f^{\text{NS}} &= -\frac{2\pi\as^2}{81}\left[4\delta_{f,u}-\delta_{f,d} \right] \, \overline{x^{-1}} \,, \label{val_qk_bfNS} \\
    {\widetilde b} &= - \frac{11\pi\as^2}{9} \, \overline{x^{-1}} \,, \label{val_qk_bt} \\
    b_2 &= \frac{2\pi\as^2}{9}\, \overline{x^{-1}}\,,\label{val_qk_b2}
\end{align}
\end{subequations}
where $\overline{x^{-1}}$ is the expectation value of the reciprocal of the longitudinal momentum fraction for a valence quark inside the proton. Its exact value depends on the applied model for the valence quark wave function, for which we use the Harmonic oscillator model from Ref.~\cite{Brodsky:1994fz} which gives $\overline{x^{-1}} = 3.64$.  We take the polarized dipole amplitudes with these parameters to correspond to initial $x=x_0=0.1$. The initial $x_0$ is chosen to be the same as in the previous global analysis~\cite{Adamiak:2023yhz}.

In Eqs.~\eqref{val_qk_params}, each parameter is fixed to be a known factor multiplied by $\as^2$, with $\as$ being the coupling constant employed in the moderate-energy calculation within the valence quark picture. Note that it needs not to follow the same prescription as $\as$ in the small-$x$ evolution. In this work, we take all the coupling constants in Eqs.~\eqref{val_qk_params}, and in the small-$x$ helicity evolution, 
to follow the coordinate-space prescription of the transverse dipole size, that is, 
\begin{align}\label{as_r}
    \as(r_{\perp}) &= \frac{12\pi}{(33-2\nf)\,\ln\left(\frac{4C^2}{r^2_{\perp}\Lambda^2_{\text{QCD}}}\right)} \, ,
\end{align}
where $\Lambda_{\text{QCD}}=\SI{0.241}{GeV}$ is the soft QCD scale and $C$ is a parameter that will be discussed in detail below. Specifically, we employ the daughter-dipole prescription for the evolution equation, following the choice in~\cite{Adamiak:2023yhz}.

In phenomenological applications where the initial condition for the unpolarized dipole amplitude evolved by the BK equation is fitted to DIS structure function data, the parameter, $C$, in Eq.~\eqref{as_r} is typically taken to be a free parameter~\cite{Albacete:2010sy}. A generic estimate connecting the coordinate and momentum space running couplings is $C^2=e^{-2\gamma_E}$~\cite{Kovchegov:2006vj,Lappi:2012vw}. Upon performing the fit with 10 different values of $C^2$ ranging from 0.05 to 2, we have found that the available polarized DIS and SIDIS data do not constrain $C^2$ accurately, and an equally good fit can be achieved with $0.25<C^2< 1.1$. Consequently, in this work, we use both the theoretically motivated value, $C^2=e^{-2\gamma_E}\approx 0.3$, and  $C^2=1$. The second case is included in order to quantify the sensitivity of our results on this scale choice, and to enable direct comparisons with the previous global analsysis~\cite{Adamiak:2023yhz} where $C^2=1$ is also used.



With all the ingredients specified by the valence quark model, only 8 parameters in Eqs.~\eqref{Born_IC} -- $c_u$, $c_d$, $c_s$, $c_u^{\text{NS}}$, $c_d^{\text{NS}}$, $c_s^{\text{NS}}$, ${\widetilde c}$ and $c_2$ -- remain to be fixed via the fit to the polarized DIS ans SIDIS data. The results of the fit are detailed in the next section.

\section{Results}\label{sec:results}

The remaining free parameters not constrained by the perturbative valence-quark model calculation are determined by performing a fit to polarized DIS data. This includes 226 data points of polarized DIS and SIDIS measurements  within the kinematic range $0.005\leq x\leq 0.1$ and $1.69\text{ GeV}^2\leq Q^2\leq 10.4\text{ GeV}^2$ from SLAC~\cite{E142:1996thl,E143:1998hbs,E154:1997xfa,E155:1999pwm,E155:2000qdr}, EMC~\cite{EuropeanMuon:1989yki}, SMC~\cite{SpinMuon:1997yns,SpinMuon:1998eqa,SpinMuon:1999udj}, COMPASS~\cite{COMPASS:2009kiy,COMPASS:2010hwr,COMPASS:2010wkz,COMPASS:2015mhb,COMPASS:2016jwv} and HERMES~\cite{HERMES:1997hjr,HERMES:1999uyx,HERMES:2004zsh,HERMES:2006jyl} for the longitudinal double spin asymmetry $A_1$, $A_{||}$ and $A_1^h$. These asymmetries probes the difference between the lepton-target cross sections where the lepton and target spins are parallel or antiparallel. Explicit definitions are given in Ref.~\cite{Adamiak:2023yhz}, together with their relations with the polarized dipole amplitudes.
The fit setup is almost identical to that of Ref.~\cite{Adamiak:2023yhz} where a good description of the data was obtained using a more flexible initial condition given in Eqs.~\eqref{Born_IC} with 24 free parameters. The differences (that have a numerically negligible effect) are outlined in the Supplementary Material.



As discussed in Sec.~\ref{sec:setup}, the valence quark model calculation fixes 16 of the 24 parameters considered to be free in Ref.~\cite{Adamiak:2023yhz}. However, we still obtain a good description of the world data with only the 8 remaining free parameters. Overall, we get $\chi^2/N_{\text{pts}} = 1.26$ with $C^2=1$ and $\chi^2/N_{\text{pts}} = 1.28$ with $C^2=e^{-2\gamma_E}$. 
This is slightly worse than $\chi^2/N_{\text{pts}}=1.03$ obtained in Ref.~\cite{Adamiak:2023yhz}, but we emphasize that the number of free parameters in our case is smaller by a factor of 3. The comparison plots to both polarized DIS and SIDIS data are shown in the Supplementary Material, illustrating an overall good agreement. 
The best-fit values for the model parameters are shown in Table.~\ref{table:params}. Parametrizations for Monte Carlo replicas, which enable one to propagate uncertainty to other observables sensitive to polarized dipole amplitudes, are also available as Supplementary Material.


\begin{table*}[tb]
    \centering
        \begin{tabular}{c|c|c|c|c|c|c|c|c||c}
       $C^2$ & $c_u$ & $c_d$  & $c_s$ & $c_u^\mathrm{NS}$  & $c_d^\mathrm{NS}$  & $c_s^\mathrm{NS}$ &  ${\widetilde c}$ & $c_2$ & $\chi^2/N_{\text{pts}}$\\
        \hline
        1 & -62.7$\pm$ 8.1 & 36.3$\pm$ 8.2 & -15.9$\pm$ 17.1 & 19.3$\pm$ 0.6 & -6.4$\pm$ 1.3 & 
-0.7$\pm$ 0.9 & 4.0$\pm$  8.2& -4.2$\pm$ 3.2 & 1.26
        \\
        \hline
       $e^{-2\gamma_E}$ &-37.5$\pm$5.5 & 27.4$\pm$ 4.7 & -8.1 $\pm$ 10.6 & 19.5 $\pm$ 0.6 & -6.5$\pm$ 1.3 & -0.6$\pm$ 0.8 & -12.3 $\pm$ 4.1 & -5.7$\pm$ 2.2 & 1.28\\
    \end{tabular}
    \caption{The best fit values for the parameters in the initial condition~\eqref{Born_IC} of the polarized dipole amplitude, along with the 1-standard-deviation uncertainty in the distribution of replicas. Recall that most of the parameters are fixed via the valence quark model to follow Eqs.~\eqref{val_qk_params}.}
    \label{table:params}
\end{table*}

Before we proceed, it is worth noting that $\chi^2/N_{\text{pts}}$ of 1.26 and 1.28 obtained in this work is significantly lower than the one obtained when the valence quark model was employed to describe the moderate-$x$ initial condition for the unpolarized DIS process under similar setting~\cite{Dumitru:2020gla,Dumitru:2021tvw,Dumitru:2023sjd}. There, the scattering amplitude was written in the dipole picture and evolved to small $x$ via the BK evolution equation. With the valence quark model constraining the initial conditions, the resulting fit yielded $\chi^2/N_{\text{pts}}=2.3$ from a DIS data set with 38 data points~\cite{Dumitru:2023sjd}. We believe that the main reason why the valence quark model results in a much better fit for helicity-dependent DIS is that the dominant contribution to the small-$x$ helicity evolution is within the DLA, resumming $\as\ln^2(1/x)$ per step of evolution~\cite{Kovchegov:2015pbl,Kovchegov:2016zex,Kovchegov:2018znm,Cougoulic:2022gbk}. In contrast, as discussed above, the BK evolution equation is single-logarithmic, resumming $\as\ln(1/x)$ per step of evolution~\cite{Balitsky:1995ub,Kovchegov:1999yj}.
Additionally, the unpolarized data is also significantly more precise, which amplifies the differences in the numerical values obtained for $\chi^2/N_{\text{pts}}$.
As a result, the small-$x$ helicity evolution becomes significant already at $x_0\sim 0.1$~\cite{Adamiak:2023yhz}, which is roughly the regime of applicability for the valence quark model~\cite{Dumitru:2024pcv,Dumitru:2020gla}. This upper limit in $x$ for the helicity evolution to kick in is much larger than that of the unpolarized BK counterpart that only becomes significant at $x\lesssim 0.01$~\cite{Albacete:2010sy}. Hence, it is expected that the constraints from the valence quark model would result in a better match to experimental data in the helicity-dependent setting.

Next, we study the proton structure function $g_1^p$ that quantifies the difference in the quark distributions between the states where the proton helicity is aligned or anti-aligned with that of the virtual photon in a polarized DIS process. At asymptotically small $x$, the helicity evolution results in $g_1^p$ growing as a power law, $x^{-\alpha_h}$ (neglecting the running coupling effects). Although this asymptotic behavior is not realized in the $x$ range covered by the available data, the evolution will eventually drive $g_1^p$ to large positive or negative values at sufficiently small~$x$. However, the fit from Ref.~\cite{Adamiak:2023yhz} is not able to distinguish the positive- and negative-$g_1^p$ scenarios. 



\begin{figure}
    \centering
    \includegraphics[width=\columnwidth]{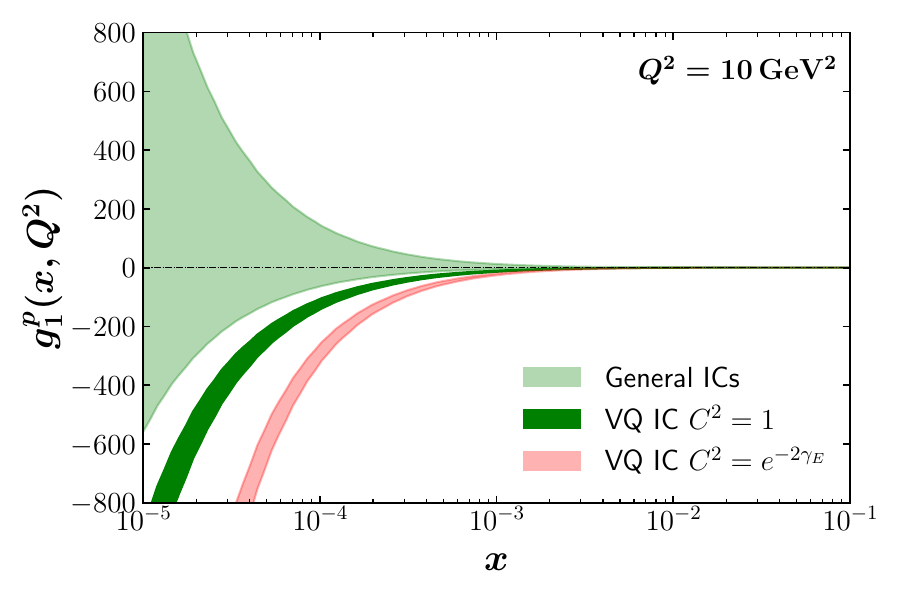}
    \caption{The resulting $g_1$ structure function for the proton, together with the $1\sigma$ uncertainty bands, calculated from the fit of small-$x$ helicity evolution to the polarized DIS and SIDIS measurements. The dark green and pink bands correspond to the initial condition with constraints from the valence quark model using the coordinate-space running coupling with $C^2=1$ and $C^2=e^{-2\gamma_E}$, respectively. The light green band corresponds to the Born-inspired initial condition employed in~\cite{Adamiak:2023yhz}.}
    \label{fig:g1p}
\end{figure}

The obtained proton $g_1^p$ structure functions are shown in Fig.~\ref{fig:g1p}. 
Specifically, we calculate $g_1^p$ as a function $x$ at typical $Q^2=\SI{10}{\;GeV^2}$ using our initial conditions with the constraints from the valence quark model, while employing the coordinate-space running coupling prescriptions with $C^2=1$ and $C^2=e^{-2\gamma_E}$, c.f. Eq.~\eqref{as_r}. The obtained $g_1^p$'s are respectively referred to as ``VQ  IC $C^2=1$'' and ``VQ  IC $C^2=e^{-2\gamma_E}$''.
For comparison, the result of the fit of Ref.~\cite{Adamiak:2023yhz}, employing the more flexible initial condition given in Eqs.~\eqref{Born_IC}, is shown in Fig.~\ref{fig:g1p} and referred to as ``General ICs''.
The input from the valence quark model appears to rule out the positive-$g_1^p$ scenario at asymptotically small $x$.
The valence quark model constrained fits are found to definitely predict $g_1^p<0$ at $x\lesssim 10^{-3}$, unlike the previous analysis without these additional constraints. However, the results from the two valence-quark fits are outside of each other's $1\sigma$ uncertainty band, but given the large uncertainty of the General IC fit this difference is not very significant.
The two running coupling scale choices result in the same negative sign for the $g_1^p$ at small-$x$, but the predicted magnitudes differ significantly at $x\lesssim 10^{-3}$, alhtough an almost identical description of the available polarized DIS data is achieved. We however note that this is the region where there are currently no experimental data available, and future measurements at the next-generation Electron-Ion Collider~\cite{Accardi:2012qut} will be able to provide further constaraints.

Next we calculate the contribution to the proton spin from the small-$x$ region. We emphasize that the applied framework does not properly include high-$x$ degrees of freedom and as such the total proton spin can not be obtained from the applied small-$x$ calculation. We compute the truncated moment of the contribution to the proton spin from the small-$x$ flavor singlet sector up to $x=0.1$:
\begin{align}
    \Delta\Sigma_{[x_\mathrm{min}]} = \int_{x_\mathrm{min}}^{0.1} \dd x \left[ \Delta u^+(x,Q^2)  + \Delta d^+(x,Q^2) \right. \\
    \left. + \Delta s^+ (x,Q^2)\right]. \notag
\end{align}
Recall that $x_0=0.1$ is the largest value of $x$ at which the double-logarithmic small-$x$ helicity evolution applies; it is also the approximate regime of the valence quark picture. Here, $\Delta u^+,\Delta d^+$ and $\Delta s^+$ refer to the polarized parton distribution functions that describe the net contribution to the proton spin from the given parton flavor. The gluon contribution $\Delta G$ is defined similarly.

\begin{figure*}
    \centering
    \includegraphics[width=\textwidth]{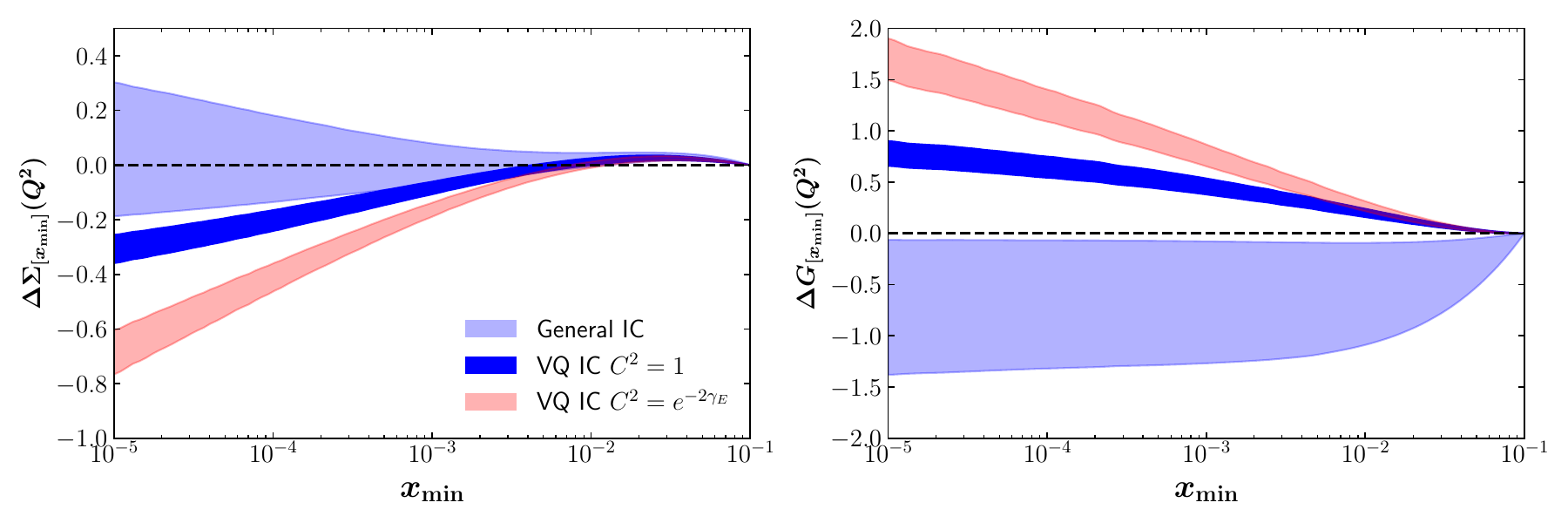}
    \caption{The resulting truncated moment of the quark-singlet (left panel) and gluon (right panel) hPDFs, together with the $1\sigma$ uncertainty bands, calculated from the fit of small-$x$ helicity evolution to the polarized DIS and SIDIS measurements. The dark blue and pink bands correspond to the initial condition with constraints from the valence quark model using the coordinate-space running coupling with $C^2=1$ and $C^2=e^{-2\gamma_E}$, respectively. The light blue band corresponds to the Born-inspired initial condition employed in~\cite{Adamiak:2023yhz}.}
    \label{fig:detalsigmadeltag}
\end{figure*}

The obtained contributions to the proton spin are shown in Figs.~\ref{fig:detalsigmadeltag}. Again, the results are compared to those obtained in Ref.~\cite{Adamiak:2023yhz} using a more flexible parametrization. As expected, the small-$x$ contribution is now much more strongly constrained. In particular, we predict a negative contribution from the quark sector, and a significantly larger positive contribution from the gluon sector, at $Q^2=10 \text{\;GeV}^2$. This is in contrast to the general IC case of Ref.~\cite{Adamiak:2023yhz}, where the quark and gluon contributions are both compatible with zero. 
Note that the sign of gluon hPDF obtained in this work is consistent with the findings from~\cite{Hunt-Smith:2024khs}.
Similarly, as in the case of $g_1^p$, the valence quark model initial condition results in $\Delta \Sigma$ and $\Delta G$ that are outside the $1\sigma$ uncertainty bands of the general IC fit from Ref.~\cite{Adamiak:2023yhz}.  

Integrating over $x$, the total small-$x$ contribution to the proton spin can be written as
\begin{equation}\label{S_small_x}
    S_{\text{small }x} = \int_{10^{-5}}^{0.1} \dd x \left( \frac{1}{2}\Delta\Sigma + \Delta G \right) ,
\end{equation}
with the upper limit being the edge of our small-$x$ regime, $x_0=0.1$. Here, in order to calculate the spin numerically, we set the lower limit to $x_{\min}=10^{-5}$, below which the non-linear, single-logarithmic contributions to the small-$x$ helicity evolution, including gluon saturation from the BK equation applied to the unpolarized dipole amplitude, should become  significant~\cite{Kovchegov:2021iyc}. Such contributions are beyond the scope of this letter. 

As defined in Eq.~\eqref{S_small_x}, the total spin contribution at small $x$ is found to be 
$S_{\text{small }x} = 0.63\pm 0.10$ for the running coupling prescription with $C^2=1$ and 
$S_{\text{small }x} = 1.35\pm 0.16$ for $C^2 = e^{-2\gamma_E}$. 
These results can be compared to $S_{\text{small }x} = -0.64 \pm 0.60$ obtained in Ref.~\cite{Adamiak:2023yhz} using the generic initial condition. Although none of the three results are compatible within the uncertainties, the former two results with the valence-quark constraints are significantly distinct from zero, that is, the total spin contribution from small $x$ is completely determined to be positive. Furthermore, it is worth noting that the total spin coming from quark or gluon helicity, as resulting from the fits in this work, varies significantly with the choice of $x_{\min}$. Ultimately, the takeaway of these results is that there is a significant amount of spin from small-$x$ partons, whose exact quantity depends on the amount and the manner the single-logarithmic effects that enter the small-$x$ helicity evolution.   


\section{Conclusion and Future Works}\label{sec:conclusion}

We have fit the initial condition of the small-$x$ helicity evolution equation, the polarized dipole amplitudes, to the available polarized DIS data. The initial condition is further constrained by the valence quark model calculation valid at moderately large $x\sim 0.1$~\cite{Dumitru:2024pcv}, where the proton non-perturbative structure is described in terms of valence quarks and constrained by form factor data, on top of which a perturbative gluon emission is included. 

These additional constraints allow us to fix most of the parameters that were left free in the previous global analysis~\cite{Adamiak:2023yhz}. Consequenlty, the remaining non-perturbative model parameters are tightly constrained, while still maintaining a good agreement with the world data. Tight constraints obtained from the fit allow us to precicesly predict the evolution of the polarized structure functions to small $x$. In particular, we find that the proton structure function $g_1^p$ becomes negative at small $x\lesssim 10^{-3}$. Furthermore, the contribution to the proton spin form small-$x$ quarks is negative, and from the small-$x$ gluons positive. Predictions for all these quantities should be compared to those of~\cite{Adamiak:2023yhz}, where  all such predictions were compatible with zero. Finally, the contribution of the quark and gluon spins at $x<0.1$ to the proton spin is found to be significant and positive.
However, as noted at the end of Section~\ref{sec:results}, the exact amount of small-$x$ spin contribution depends largely on the single-logarithmic corrections to the helicity evolution, in addition to the applied running coupling scale choice.

The best fit parametrization for the polarized dipole amplitudes, Eqs.~\eqref{Born_IC}, with the valence quark model constraints shown in Eqs.~\eqref{val_qk_params}, is shown in Table~\ref{table:params}. The comparison between the predictions from our fit and the experimental polarized DIS measurements for each asymmetry observable is provided in the Supplementary Material.

The constraints from the valence quark model can be systematically improved.
Most importantly, one could include the Melosh rotation~\cite{Melosh:1974cu,Hufner:2000jb} in the valence quark model, taking into account the fact that the helicity and the longitudinal spins of the valence quarks are not identical in the case of non-zero transverse momentum. The inclusion of Melosh rotation into our model will possibly lead to better agreement between the hPDFs resulting from the valence-quark-based fit and those from the Born-based fit~\cite{Adamiak:2023yhz}. Finally, although technically challenging, the valence quark model can in principle be systematically improved by considering higher Fock states than $|qqq\rangle$ and $|qqqg\rangle$ considered in this work and in Ref.~\cite{Dumitru:2024pcv}.

On the experimental side, the future Electron-Ion Collider (EIC) will significantly extend the kinematical domain covered by polarized DIS measurements~\cite{Accardi:2012qut}, and as such can be expected to provide strong additional constraints to the extraction of the proton spin contribution from small-$x$ quarks and gluons. We expect the EIC results not only to significantly reduce the uncertainties of proton's $g_1^p$ structure function and parton hPDFs at small $x$, but also to provide a glimpse into their behaviors at very small $x$ where the effects of single-logarithmic corrections set in.

Furthermore, given the ongoing development in calculating the spin asymmetry in forward proton-proton and proton-nucleus collisions based on the small-$x$ helicity evolution~\cite{Kovchegov:2024aus}, there should be more comprehensive fits that involve these particle production data on top of the polarized DIS and SIDIS measurements considered here and in Ref.~\cite{Adamiak:2023yhz}. This will likely provide additional constraints to the initial condition and shed light on the accuracy of the valence quark picture as the moderate-$x$ framework for the initial condition of the small-$x$ helicity evolution.

Recently, a reformulation to the helicity evolution has been proposed~\cite{Borden:2024bxa} in order to be able to express quark and gluon hPDFs in terms of the polarized dipole amplitudes at the same orders in coupling constant. This change introduces new types of polarized dipole amplitudes that would add free parameters to the initial condition of the evolution equation as a whole. As a result, the constraints from the valence quark model are expected to prove even more vital to future phenomenological analyses, as a calculation similar to~\cite{Dumitru:2024pcv} can be performed on the new dipole types from~\cite{Borden:2024bxa} as well. At the end, the total number of free parameters are expected to reduce by a factor of three compared to the Born-inspired initial condition, in a similar fashion as in this work and in~\cite{Dumitru:2024pcv}.

In the long run, the valence quark model could provide significant improvements to similar fits for small-$x$ evolution equations relevant to other transverse-momentum-dependent PDFs (TMDPDFs). Such evolution equations also contain dominant contributions that are double-logarithmic~\cite{Kovchegov:2022kyy,Santiago:2023rfl,Santiago:2024iem}, which implies that the valence quark model could provide useful constraints to their respective moderate-$x$ initial conditions. However, since the available polarized small-$x$ data relevant to these TMDs remain limited, the fits mentioned above are expected to only yield statistically significant predictions once the EIC results come out~\cite{Accardi:2012qut,Abir:2023fpo}.







\begin{acknowledgments}
    DA is supported by the U.S. Department of
Energy, Office of Science, Office of Nuclear Physics under Award Number DE-AC05-06OR23177 under
which Jefferson Science Associates, LLC, manages and operates Jefferson Lab.
    HM and YT are supported by the Research Council of Finland, the Centre of Excellence in Quark Matter and projects 338263, 346567 and 359902, and by the European Research Council (ERC, grant agreements No. ERC-2023-COG-101123801 GlueSatLight and No. ERC-2018-ADG-835105 YoctoLHC). The content of this article does not reflect the official opinion of the European Union and responsibility for the information and views expressed therein lies entirely with the authors.
\end{acknowledgments}

\bibliographystyle{JHEP-2modlong}

\providecommand{\href}[2]{#2}\begingroup\raggedright\endgroup

\pagebreak
\widetext
\begin{center}\newpage
\textbf{\large Supplementary Materials}
\end{center}
\setcounter{equation}{0}
\setcounter{figure}{0}
\setcounter{table}{0}
\setcounter{page}{1}
\makeatletter
\renewcommand{\theequation}{S\arabic{equation}}
\renewcommand{\thefigure}{S\arabic{figure}}
\renewcommand{\bibnumfmt}[1]{[S#1]}
\renewcommand{\citenumfont}[1]{S#1}

\begin{figure*}[b]
    \includegraphics[width=\textwidth]{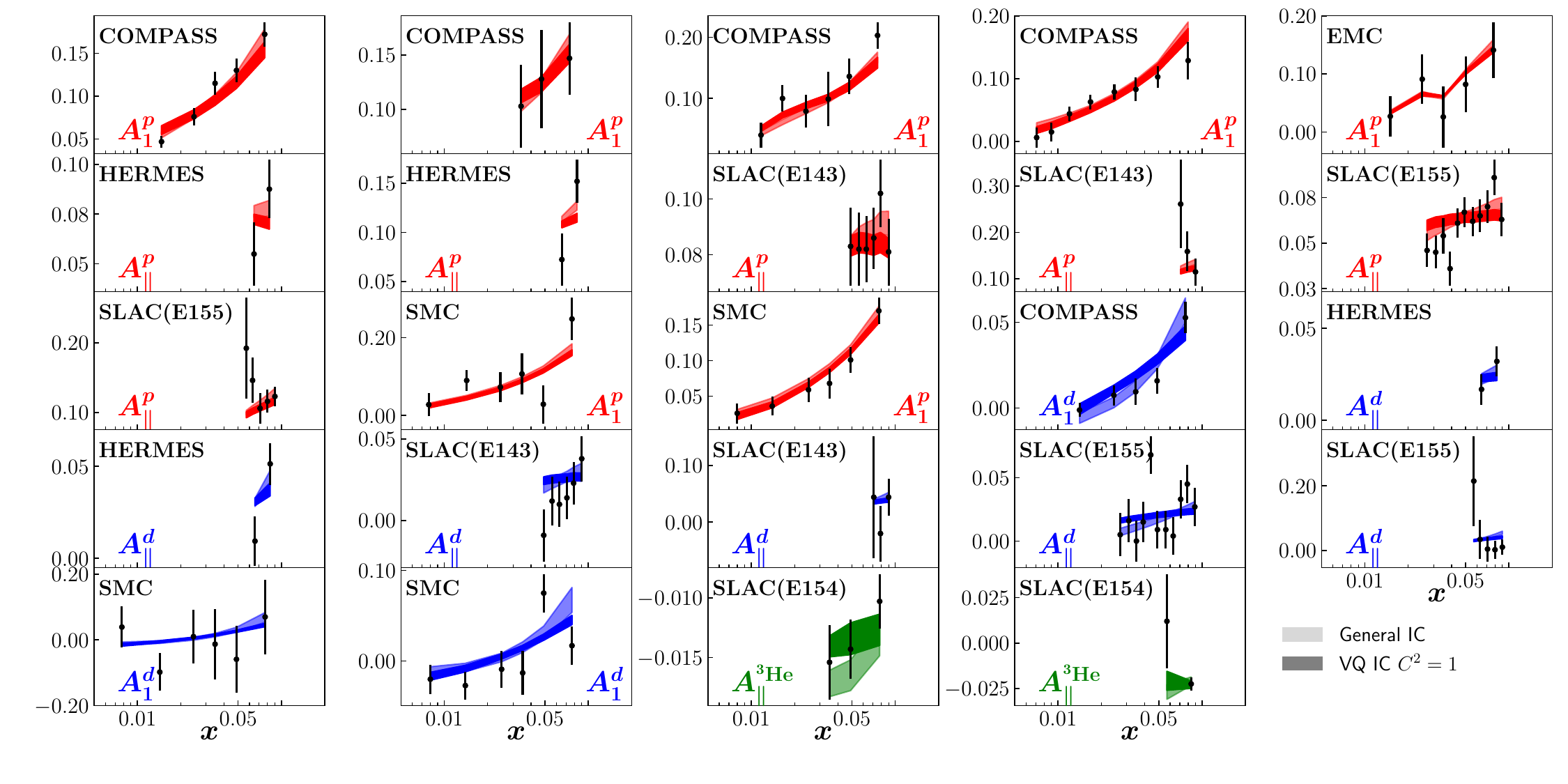}
    \caption{Comparison plots for asymmetry obervables in polarized DIS processes. Here, the coordinate-space running coupling is employed with $C^2=1$ for both initial conditions.}
    \label{fig:data_comparison_DIS}
\end{figure*}

\begin{figure*}
    \includegraphics[width=\textwidth]{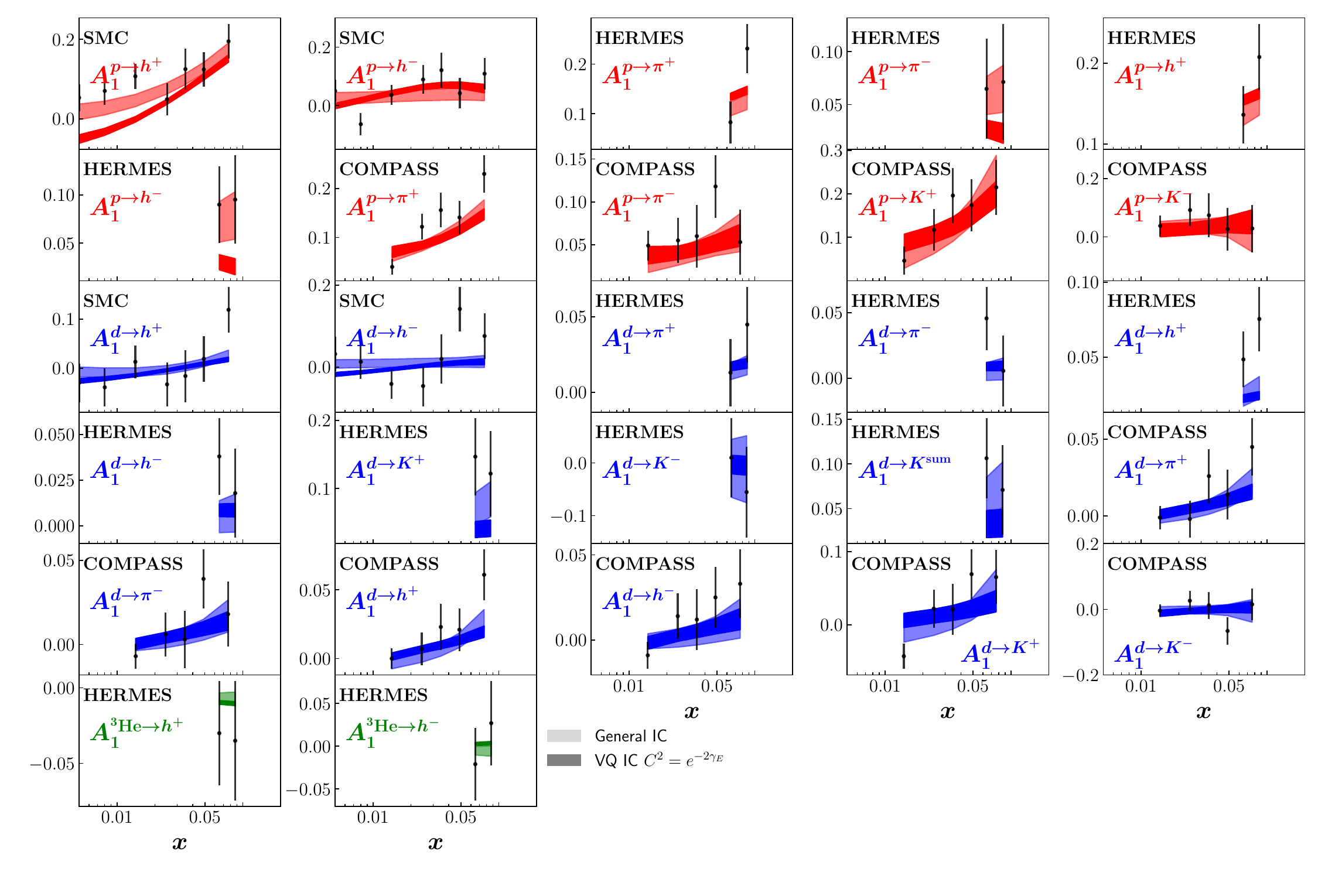}
    \caption{Comparison plots for asymmetry obervables in polarized SIDIS processes. Here, the coordinate-space running coupling is employed with $C^2=1$ for both initial conditions.}
    \label{fig:data_comparison_SIDIS}
\end{figure*}

\begin{figure*}
    \includegraphics[width=\textwidth]{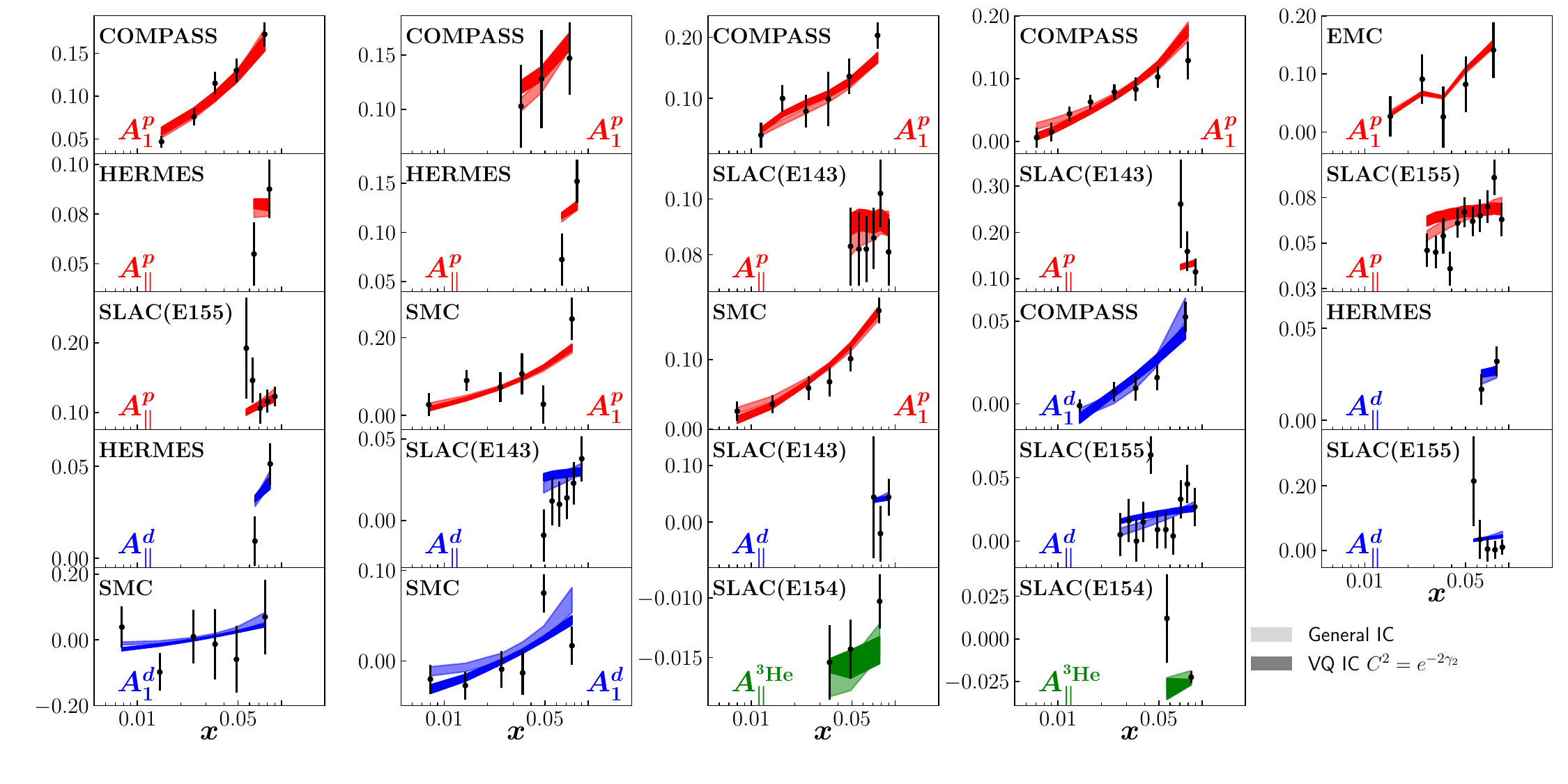}
    \caption{Comparison plots for asymmetry obervables in polarized DIS processes. Here, the coordinate-space running coupling is employed with $C^2=e^{-2\gamma_E}$ for both initial conditions.}
    \label{fig:data_comparison_DIS_2}
\end{figure*}

\begin{figure*}
    \includegraphics[width=\textwidth]{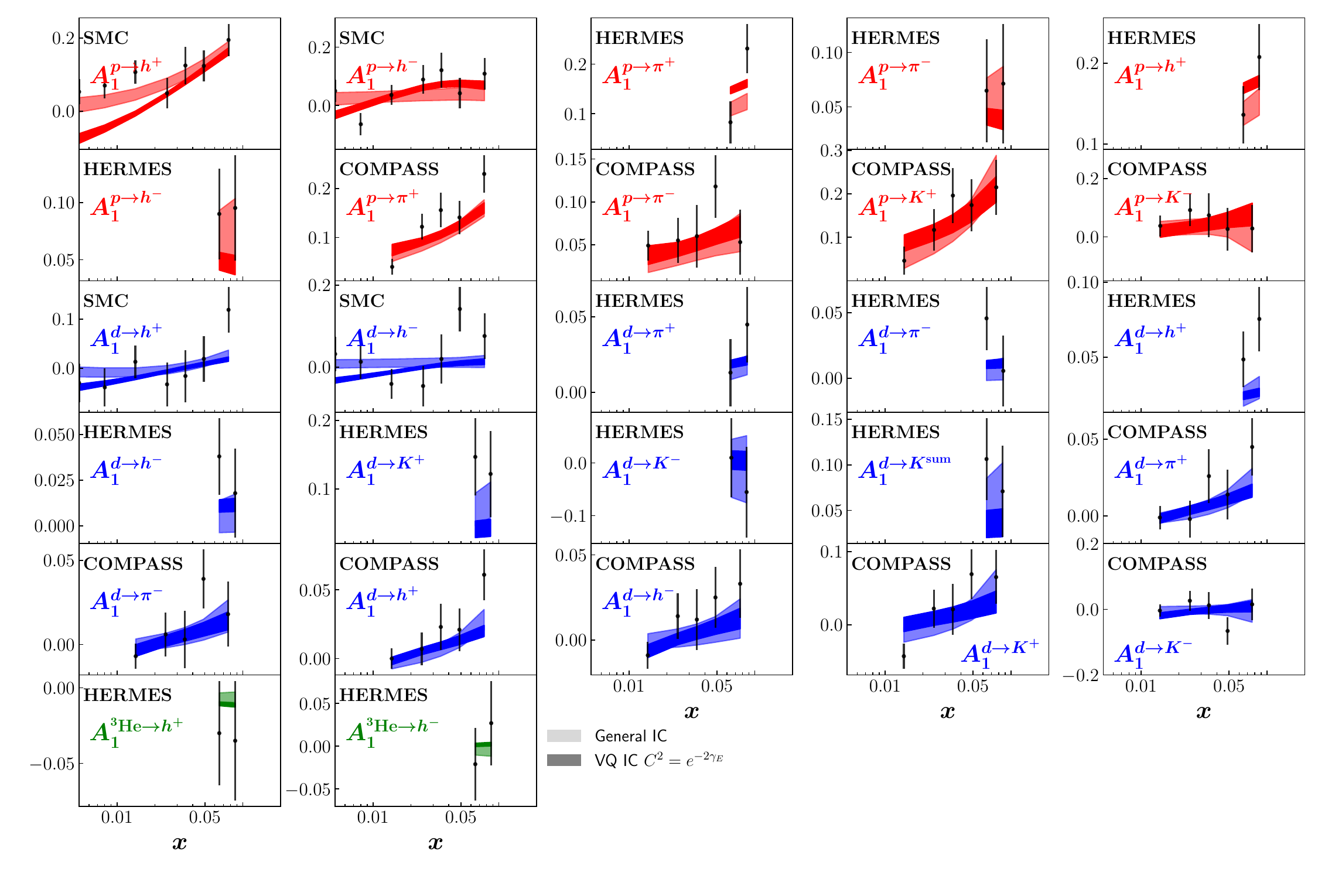}
    \caption{Comparison plots for asymmetry obervables in polarized SIDIS processes. Here, the coordinate-space running coupling is employed with $C^2=^{-2\gamma_E}$ for both initial conditions.}
    \label{fig:data_comparison_SIDIS_2}
\end{figure*}

This document displays comparison plots for the asymmetries, $A_1$ and $A_{\parallel}$, in polarized DIS and SIDIS processes, based on the fit presented in~\cite{paper}. The plots for DIS are shown in Figs.~\ref{fig:data_comparison_DIS} and \ref{fig:data_comparison_DIS_2}, for running coupling prescriptions with $C^2=1$ and $C^2=e^{-2\gamma_E}$, respectively, c.f. Eq.~(4) in~\cite{paper}, while the plots for SIDIS are displayed in Figs.~\ref{fig:data_comparison_SIDIS} and \ref{fig:data_comparison_SIDIS_2} with similar values of $C^2$. In each plot within both figures, the darker band represents the asymmetry observable calculated from the fit in~\cite{paper} with valence-quark initial condition, while the lighter band represents the same observable based on the previous fit from~\cite{Adamiak:2023yhz} with the more flexible Born-inspired initial condition, c.f. Eq.~(2) in the letter~\cite{paper}. Note that the Born-inspired initial condition, i.e. the ``General IC'' case, in each of the plots is employed with $C^2=1$.
Finally, in each plot, the experimental measurement for the labeled asymmetry observable is shown as data points together with the error bars. The experiments include 226 data points within $0.005\leq x\leq 0.1$ and $1.69\text{ GeV}^2\leq Q^2\leq 10.4\text{ GeV}^2$ from SLAC~\cite{E142:1996thl,E143:1998hbs,E154:1997xfa,E155:1999pwm,E155:2000qdr}, EMC~\cite{EuropeanMuon:1989yki}, SMC~\cite{SpinMuon:1997yns,SpinMuon:1998eqa,SpinMuon:1999udj}, COMPASS~\cite{COMPASS:2009kiy,COMPASS:2010hwr,COMPASS:2010wkz,COMPASS:2015mhb,COMPASS:2016jwv} and HERMES~\cite{HERMES:1997hjr,HERMES:1999uyx,HERMES:2004zsh,HERMES:2006jyl} for the proton, deuterium and helium-3 targets. The quality of fit is measured using the $\chi^2/N_{\text{pts}}$ metric. With the flexible, Born-inspired initial conditions, $\chi^2/N_{\text{pts}} = 1.03$. For the valence-quark model in this present work, we obtain $\chi^2/N_{\text{pts}} = 1.26$ with $C^2=1$ and $\chi^2/N_{\text{pts}} = 1.28$ with $C^2=e^{-2\gamma_E}$.

It should be noted that the results for the flexible, Born-inspired initial conditions presented here and in~\cite{paper} are not exactly the same as those presented in \cite{Adamiak:2023yhz}. As explained in \cite{Anderson:2024evk}, a mistake in the normalization of the COMPASS data sets in the Jefferson Lab Angular Momentum (JAM) database was found. The fragmentation functions and denominator of the asymmetry used in fits of this work are based on said COMPASS data and needed to be updated. We repeated the fit and analysis of~\cite{Adamiak:2023yhz} using the updated fragmentation functions and found a negligible difference.

\bibliographystyle{JHEP-2modlong}
\providecommand{\href}[2]{#2}\begingroup\raggedright\endgroup

\end{document}